# Robust Electron Pairing in the Integer Quantum Hall Effect Regime

H. K. Choi[1*], I. Sivan[1*], A. Rosenblatt[1], M. Heiblum[1], V. Umansky[1], D. Mahalu[1]

**Electron pairing is a rare phenomenon appearing only in a few unique physical systems; e.g., superconductors and Kondo-correlated quantum dots. Here, we report on an unexpected, but robust, electron 'pairing' in the integer quantum Hall effect (IQHE) regime. The pairing takes place within an interfering edge channel circulating in an electronic Fabry-Perot interferometer at a wide range of bulk filling factors, $2<\nu_B<5$. The main observations are: (a) High visibility Aharonov-Bohm conductance oscillations with magnetic flux periodicity $\Delta\varphi=\phi_0/2=h/2e$ (instead of the ubiquitous $h/e$), with $e$ the electron charge and $h$ the Planck constant; (b) An interfering quasiparticle charge $e^*\sim 2e$ - revealed by quantum shot noise measurements; and (c) Full dephasing of the $h/2e$ periodicity by induced dephasing of the adjacent edge channel (while keeping the interfering edge channel intact) – a clear realization of inter-channel entanglement. While this pairing phenomenon clearly results from inter-channel interaction, the exact mechanism that leads to $e$-$e$ attraction within a single edge channel is not clear.**

## Introduction

The Fabry-Perot interferometer (FPI) is one of several interferometers implemented in optical[1,2] as well as in electronic systems[3–6]. Its electronic version[5,6] is an excellent platform to demonstrate the wave-particle duality of electrons: interference[3] - a manifestation of wave-like behavior, alongside with current-fluctuation[7] due to particles discreteness. Moreover, the electronic version of the FPI harbors an exceedingly more complex behavior due to the interacting nature of electrons. Two distinct regimes of the FPI operating in the QHE regimes have been observed[5,6] and theoretically studied[8,9]: (*i*) The coherent Aharonov-Bohm (AB) regime, where interactions were thought to play only a minor role; and the ubiquitous (*ii*) The Coulomb dominated (CD) regime, where interactions are dominant. In the rather difficult to achieve AB regime, the conductance of the FPI obeys strictly AB interference of independent electrons[10,11], with conductance oscillation phase follows $\varphi_{AB}=2\pi AB/\phi_0$, where $A$ denotes the area of the interferometer, $B$ the applied magnetic field, and $\phi_0=h/e=41\mu m^2 G$ the magnetic flux-quantum. Alternatively, in the CD regime, conductance oscillation is due to periodic charging (and discharging) of the FPI with single electrons. Noting that the AB regime is highly desirable since it had been proposed to be a sensitive marker to fractional statistics (abelian or non-abelian) of fractionally charged quasiparticles[12–15].

[1] Braun Center for Submicron Research, Department of Condensed Matter Physics, Weizmann Institute of Science, Rehovot 76100, Israel
[*] These authors contributed equally to this work

We address the FPI in the AB regime, which is formed by two quantum point contacts (QPCs) that serve as electronic beam splitters (Fig. 1a). An chiral edge channel impinging from the left is transmitted with amplitude $\tau = t_l t_r \left(1 + \sum_{n=1}^{\infty} (r_l r_r)^n e^{in\phi_{AB}}\right)$, where $r_l$ ($r_r$) and $t_l$ ($t_r$) are the reflection and transmission amplitudes of the left (right) QPC, respectively. Each term in the above sum stands for a different number $n$ of windings that a coherent electron undergoes. Hence, the transmission through the interferometer is an oscillating function of the acquired AB phase in a single winding in the FPI:

$$(1) \quad |\tau|^2 = \frac{t_l^2 t_r^2}{\left|1 - r_l r_r \cdot e^{i\phi_{AB}}\right|^2} ,$$

To first order in the transmission coefficient of the QPCs (namely, $r_{l,r} \ll 1$), is $|\tau|^2 = t_l^2 t_r^2 (1 + r_l^2 r_r^2 \cos \phi_{AB})$. Evidently, one expects an AB oscillation period in area $\Delta A = \phi_0/B$ and in magnetic field $\Delta B = \phi_0/A$.

As described in some detail below, the predicted behavior of Eq. 1 was precisely observed in a range of bulk filling factors $\nu_B$=1-2; however, at fillings around $\nu_B$=3-4 the AB periodicity halved, namely, $\varphi_{AB} = 2\pi A \cdot B/\phi_0^*$, with $\phi_0^* = \frac{h}{2e}$, without an any observable sign of $\phi_0$ periodicity. Since similar visibilities (contrast, defined as the oscillation amplitude divided by its average), reaching ~60%, were observed in the two filling regimes, interference of only even windings as the cause of the halved periodicity can be excluded. Testing the possibility of electron pairing in the interfering channel, the interfering charge $e^*$ was determined by measuring the quantum shot noise, indeed leading to $e^* \sim 2e$. These observations were found to be robust and reproducible in FPIs of very different sizes (2-12μm$^2$); fabricated in different heterostructures (2DEG density ranged 1.3-2.4×10$^{11}$cm$^{-2}$); and at a temperature range 30-130mK. Moreover, two very different screening methods were employed in order to suppress the CD regime and operate the FPI in the AB regime; both leading to similar results.

**Measurement setup**

Our FPIs were realized in a ubiquitous high mobility 2DEG embedded in AlGaAs-GaAs heterostructure(s), employing optical and E-beam lithography. The FPIs were fabricated as a semi-closed stadium (Fig. 1), terminated with two QPCs; one at the source side and the other at the drain side. A charged 'modulation gate' (MG) allowed varying the FPI's area. Only the most outer edge channel was partitioned by the QPCs, while the impinging inner ones were fully reflected from the QPCs and, likely, the trapped ones circulated inside the FPI. Coulomb interactions within a single channel (intra-edge interactions) were suppressed via two different configurations (illustrated in Figs. 1b & 1c); both leading to similar results: (*i*) Metallic (Ti-Au) top-gate covering nearly the whole area of the FPI; (*ii*) Small, grounded, ohmic-contact (Ni-Ge-Au), alloyed in the center of the incompressible bulk of the FPI. While the screening process achieved by the ohmic contact is less transparent, it was found to be more effective in suppressing interaction, allowing fabricating smaller FPIs operating in the AB regime. Differential source-drain conductance was measured by applying an AC voltage of 1μVrms @ 800kHz; while excess quantum shot noise at the drain was measured by driving a variable DC source current, and measuring the noise at a center frequency of 800kHz and bandwidth of 10kHz.

**Conductance measurements**

The conductance of the FPI was measured as function of the magnetic flux φ, either by varying the magnetic field or the enclosed area (via MG). In the following we present data taken with two FPIs, each with a center-ohmic contact, and areas ~2μm$^2$ and ~12μm$^2$. Figure 2 presents the most important data of this work. Starting with the smallest FPI, we plot in Fig. 2a the flux-dependent conductance at $v_B \cong 2$ (being representative of similar results in $v_B \cong 1$-2.5). Constant phase lines correspond to constant flux; namely, increasing magnetic field necessitates reducing the FPI area. An AB area of 2.1μm$^2$ is deduced from $\Delta B$ period. However, at $v_B \cong 3$ (being a representative of $v_B \cong 2.5$-4.5), an unexpected halving of the oscillation periods in $B$ and in $V_{MG}$ is observed (Fig. 2b). This is clearly evident in the $B$ dependence oscillations (at constant $V_{MG}$) shown in Fig. 2c, and their Fourier transform in Fig. 2d. Higher harmonics, corresponding to multiple windings, are also visible in the two regimes. Note an absence of any sign of the 'fundamental' periodicity $h/e$ in the $h/2e$ regime.

The universality of the results is presented in Fig. 3, this time with the larger FPI ($A$~12μm$^2$), where the dependence of the AB frequencies ($\phi_0/\Delta B$ in Fig. 3a; $1/\Delta V_{MG}$ in Fig. 3b) is plotted as function of bulk filling $v_B$. In filling ranges $v_B<2.5$ and $v_B>4.5$ the normalized oscillation frequency in magnetic field is $\phi_0/\Delta B^e=12$μm$^2$ - agreeing with FPI area, and is independent of the filling factor. The frequency in MG $1/\Delta V_{MG}$ depends linearly on $B$, $1/\Delta V_{MG}^e = \alpha B/\phi_0$, as expected. This dependence can be easily understood by recalling that the incremental change in charge δ$Q$ is related to δ$V_{MG}$ via the capacitance $C_{MG} = \frac{\delta Q}{\delta V_{MG}}$. Moreover, its relation to the incremental change in area is δ$Q$=$en_e$δ$A$, with $n_e$ the carrier density. Hence, $\alpha = \frac{C_{MG}}{en_e\left(\frac{e*}{e}\right)} = 0.17 \frac{\mu m^2}{V}\left(\frac{e*}{e}\right)$, with $e^*$ is the interfering quasiparticle charge in the most outer channel, being in this filling range $e$. However, in the filling range $v_B$=3-4.5, the $B$ dependent normalized frequency is doubled, $\phi_0/\Delta B^{2e}=24$μm$^2$, and again independent of filling factor. This doubling suggests halved flux quantum, namely, $\frac{\phi_0}{\phi_0^*} = \frac{e^*}{e} = 2$. Similarly, the AB frequency in $1/\Delta V_{MG}$ has now twice the slope, $0.34 \frac{\mu m^2}{V}$, again leading to $\frac{e^*}{e} = 2$. The small deviation from an exact $\frac{e^*}{e} = 2$ in the region $v_B$~2.5-3 (Figs. 3a and 3b) is observed in all the tested samples. The behavior in the transition region near $v_B$~2.5 and $v_B$~4.5 is not universal and will not be discussed here.

**Can $h/2e$ periodicity be attributed to even windings?**

*(a) Analysis of the visibility*

While the doubling of the AB frequencies can be attributed to $\varphi_{AB} = 2\pi A \cdot B/\phi_0^*$, with $\phi_0^* = h/e^* = h/2e$, one can also envision a peculiar preference of only even number of interfering windings; namely, $A^*=2A$. In order to distinguish between these two scenarios we address the visibility of the $h/2e$ oscillation compared to that of the $h/e$ oscillation. With the visibility of the $h/e$ periodicity in the ~12μm² FPI being $v^{(e,1)}{}_{max}\approx 60\%$, we estimated a dephasing length $L_\gamma \approx 35$μm (see Supp. 2). This value sets an upper-bound for the highest visibility of the second harmonic (two windings in the $h/e$ regime) $v^{(e,2)}{}_{max}\approx 10\%$; which had been confirmed experimentally. However, the highest visibility in the $h/2e$ regime was found to be $v^{(2e,1)}{}_{max}\approx 50\%$; refuting a peculiar favoring of even number windings.

*(b) Charge determination via quantum shot noise measurements*

Shot noise measurements were found to provide an excellent method for a determination of the partitioned charge[16–21]. For independent and stochastic partitioning events by any scatterer, the expected 'zero frequency' ($\hbar\omega<<eV_{SD}$) excess noise (added noise due to driven current) is given by Eq. 2, which had been modified to allow $e^*\neq e$[19,20]:

$$(2) \quad S_0(t,I) = 2e^* I \cdot t \left(1 - \frac{e}{e^*}t\right) \cdot F(eV_{SD}, k_B T),$$

where $t=|\tau|^2$ is the transmission coefficient, $I$ the impinging current, $T$ the electron-temperature, and $V_{SD}$ source-drain DC-bias.

Noise measurements were first conducted with a single QPC partitioning the outer edge channel at $v_B=2$ and $v_B=3$. A partitioned charge $e^*=e$ was extracted in a wide range of transmission coefficients using Eq. 2 (Figs. 4a & 4b). The FPI was then formed by pinching the second QPC with noise measured in the $h/e$ regime; leading to $e^*=e$ (Fig. 4c and Supp. 8). In the $h/2e$ regime, on the other hand, the quasiparticle charge was found consistently to be $e^*\sim 2e$ (Fig. 4d). Note that very large Fano factors were found when the visibility was very high, which is attributed to phase noise due to charge fluctuations[22,23].

**Inter-edge interaction**

To test the effect of inter-channel interaction, we modified further the FPI by adding an additional 'center-QPC' between the edge of the FPI and the center-ohmic contact (see SEM image in Fig. 5a). This configuration allows selective reflection of in-bound edge channels to the center-contact; thus dephasing them and fixing their Fermi energy at ground. Starting at filling $v_B=2$ with oscillation periodicity $h/e$, we plotted the transmission of the center-QPC (to be distinguished from the reflection to the center-ohmic) and the oscillation visibility of the most outer edge channel as function of the center-QPC gate voltage (Fig. 5b). As expected, the visibility quenches when the interfering edge channel is fully reflected to the center-ohmic contact.

Now, in $\nu_B$~3, with oscillation periodicity $h/2e$, the behavior is entirely different. The visibility of the $h/2e$ periodicity (with all its harmonics) quenches rapidly as soon as the adjacent second edge channel is reflected to the center-contact - with no sign of interference at all (Fig. 5c). We wish to stress that when the $h/2e$ periodicity quenches the outer edge channel is uninterruptedly fully transmitted through the center-QPC. Since no tunneling was observed between these two adjacent edge channels (see Supp. 9), this result demonstrates that the coherence of the second edge channel is indispensable for the observation of $h/2e$ periodicity of the outer edge channel.

Employing charge measurement throughout the above described dephasing process, we plot again in an expanded scale the visibility and the FPI conductance (proportional to the transmission of the outer edge channel) as function of $V_{center-QPC}$ in Fig. 6a, alongside with the partitioned charge in Fig. 6b. Since the conductance maintains its slope as the second channel is gradually reflected to ground, it is clear that no current from the outer channel is being lost to the second channel. However, as the coherence of the $h/2e$ periodicity gradually diminishes, the quasiparticle charge drops gradually from $e^*$~$2e$ to $e^*$=$e$. Noting, that since the transmission of the FPI is highly sensitive to those of the two QPCs and the AB phase, each charge value was determined by averaging the charge measured in a wide span of AB phases at a fixed setting of QPCs. It is hence clear that an interfering charge $e^*$~$2e$ is always associated not only with the doubling of the AB frequency, but also with a significant visibility (coherence).

**Discussion & Summary**

Observation of $h/2e$ periodicity (and higher harmonics) in other electronic systems, such as a Coulomb dominated FPI[24] and a Coulomb dominated quantum anti-dot[25–27], have been reported in the past. In these examples, the non-linear conductance exhibited the ubiquitous Coulomb diamonds[25,28], and the oscillation periodicity scaled inversely with the number of fully transmitted channels through the device[5,29–32]. These two features clearly differ from our observation in the AB - FPI.

It may be useful to recapitulate the main results in this work for the interference of the most outer edge channel of the FPI: (*i*) While the 'bare FPI' is heavily dominated by Coulomb interactions, thus fully masking the AB interference, our FPI had been modified to suppress the interactions (mostly the intra-cannels) and thus showed clear AB behavior. Two methods were implemented for screening the FPI: (a) An added top gate; (b) A grounded small center-ohmic contact to the bulk. (*ii*) Different area interferometers, ranging from ~2μm² to ~12μm², fabricated on different heterostructures, showed qualitatively similar results. (*iii*) The FPIs were tested in a wide range of QPCs transmissions. (*iv*) At bulk fillings $\nu_B$=1-2.5 the AB periodicity in magnetic field and in modulation gate voltage (affecting the AB area) corresponded to flux periodicity of $\phi_0$=$h/e$; namely, single electron interference. At this range the interfering charge, determined by shot noise measurement, was $e^*$=$e$. (*v*) At bulk fillings $\nu_B$=2.5-4.5, the periodicity in magnetic field and modulation gate voltage halved; namely, with a flux periodicity $\phi_0$=$h/2e$. The interfering charge, measured by shot noise, was $e^*=2e$. (*vi*) Dephasing the second edge channel (second spin-split Landau level) in filling range $v_B$=1-2.5 (by shorting the second channel to the center-ohmic) did not affect the *h*/*e* oscillation and the interfering charge. (*vii*) Dephasing the second edge channel in filling range $v_B$=2.5-4.5 fully dephased the *h*/2*e* oscillation and simultaneously lowered the partitioned charge to $e^*=e$. (*viii*) Tunneling current between adjacent edge channels was not observed (see also Supp. 9). (*ix*) The temperature dependence of the visibility in the *h*/*e* and the *h*/2*e* regimes was very similar (see Supp. 4); decaying exponentially $e^{-T/T_0}$, with $T_0$~30mK. (*x*) The dependence of the visibility on the transmission of the QPCs was very similar in both regimes (see Supp. 3). (*xi*) While the transition regions near $v_B$=2.5 and $v_B$=4.5 are complex and sample dependent, near $v_B$=4.5 a coexistence of the two periods, *h*/*e* and the *h*/2*e* - being in phase - was observed (see Supp. 5). Moreover, dephasing the *h*/2*e* periodicity (as in (*vi*)) did not affect the *h*/*e* periodicity (see Supp. 5). (*xii*) Allowing the second edge channel to interfere (while the outer channel is fully transmitted), we find only the ubiquitous *h*/*e* oscillation in all bulk filling factors (see Supp. 3).

Our results reveal an emergent, and robust, electron pairing in a coherent chiral edge channel, hence intimately tied to an interfering process. Clear evidence of inter-channel entanglement between the interfering channel and the adjacent one takes place under the 'pairing' conditions. While Cooper pairing is phonon mediated, here the exact mechanism that leads to intra-channel two-electron attraction is not understood. An important question remains whether the observed phenomenon is general and can be reproduced in other correlated quantum systems.

**Acknowledgements** We thank Y. Gefen, W. R. Lee and Y. Cohen for useful discussions and ideas. We also thank a numerous number of colleagues we consulted with in the effort to understand the mechanism that governs the observed effects. We acknowledge the partial support of the Israeli Science Foundation (ISF), the Minerva foundation, the U.S.-Israel Bi-National Science Foundation (BSF), and the European Research Council under the European Community's Seventh Framework Program (FP7/2007-2013)/ERC Grant agreement No. 227716.

**Author Contributions** H. K. C., I. S. and M. H. contributed to paper writing. H. K. C., I. S. and A. R. contributed to sample design, device fabrication and data acquisition. D. M. contributed to electron beam lithography. V. U. grew the heterostructures.

**Author information**. Correspondence and requests for materials should be addressed to M.H. (moty.heiblum@weizmann.ac.il)

**Figure 1**

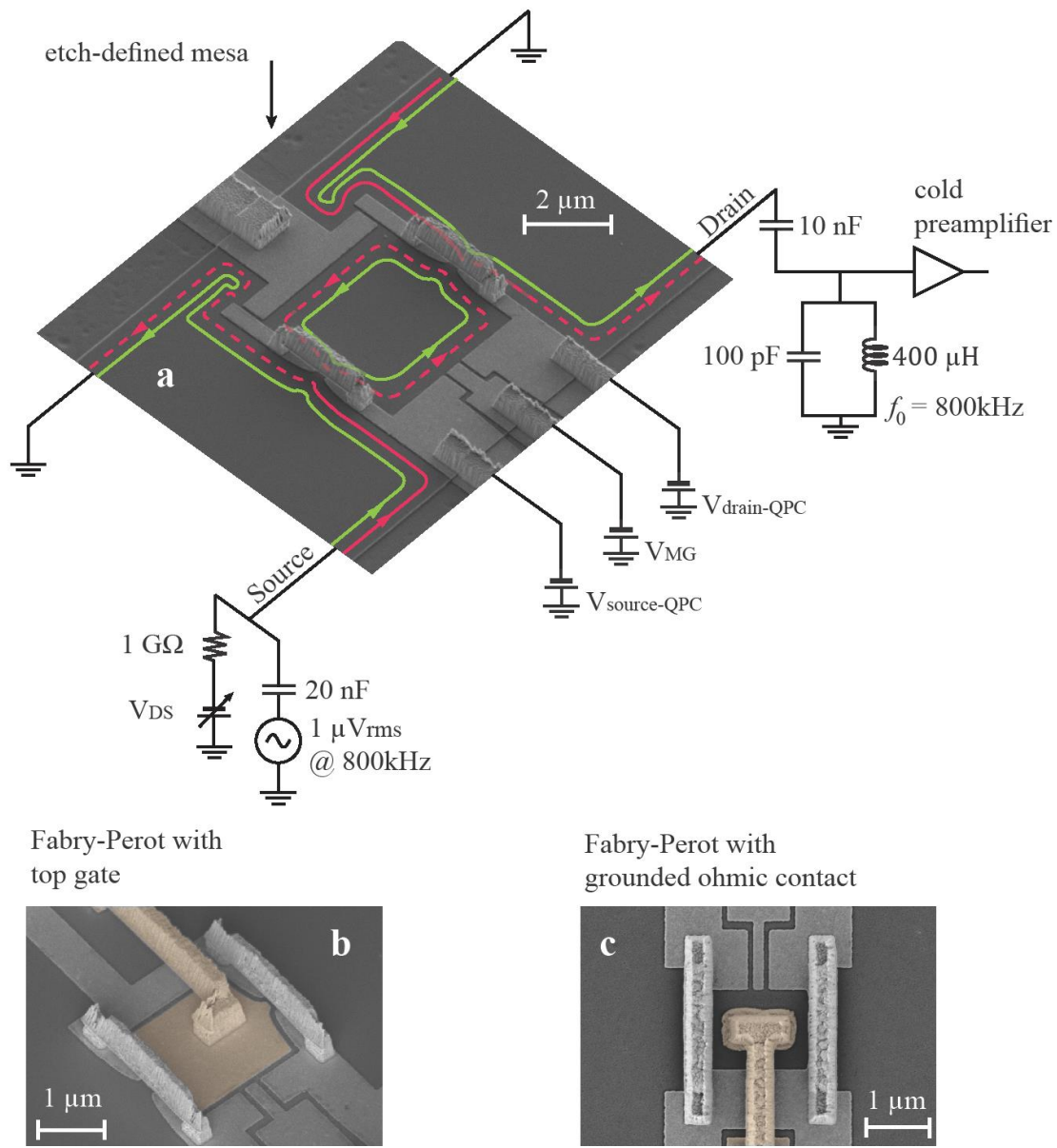

**Figure 1 | SEM images of Fabry-Perot interferometers and the experimental set-up. a**, SEM image of a FPI with an illustration of chiral edge channels at $v_B$=2. Edge-channels run along the interfaces between the mesa (regions with a 2DEG) and regions depleted by etching or gating. The Hall bar is defined by etching and the FPI by gates. Air bridges are used to connect gates and the ohmic contact to the relevant potentials. b, SEM image of Top-gated FPI. c, SEM image of a FPI in center-ohmic contact. Different FPI sizes were used in the experiments.

**Figure 2**

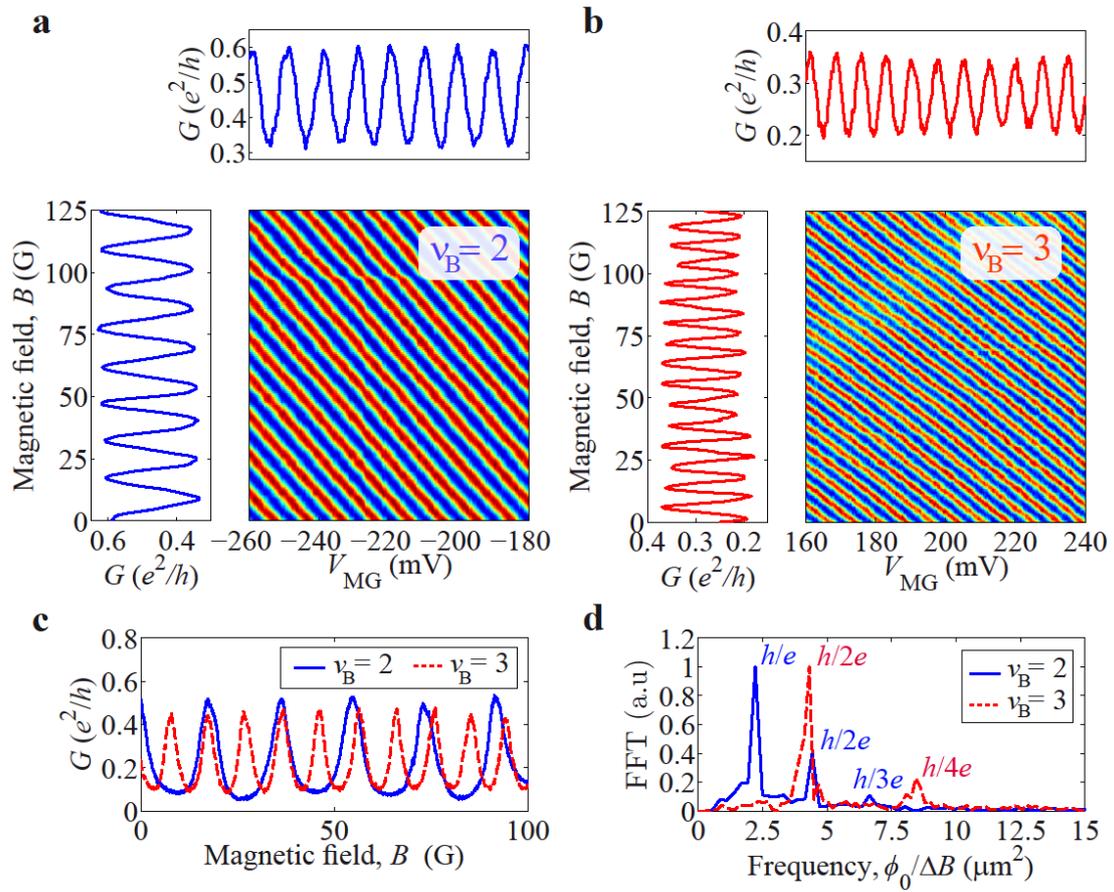

**Figure 2 | Aharonov-Bohm interference in the *h/e* and *h/2e* regimes measured with a 2μm$^2$ FPI with grounded center-ohmic contact. a**, **b**, Conductance G of the FPI versus both magnetic field *B* and modulation-gate voltage $V_{MG}$ in the *h/e* regime **a,** and in the *h/2e* regime **b,** measured at $\nu_B$~2 & 3 respectively. **c**, Characteristic AB oscillations with respect to magnetic-field *B* in the two regimes. **d**, Corresponding Fourier transforms; noticeably, the second harmonic of the *h/e* periodicity coincides with the first harmonic of *h/2e* periodicity.

**Figure 3**

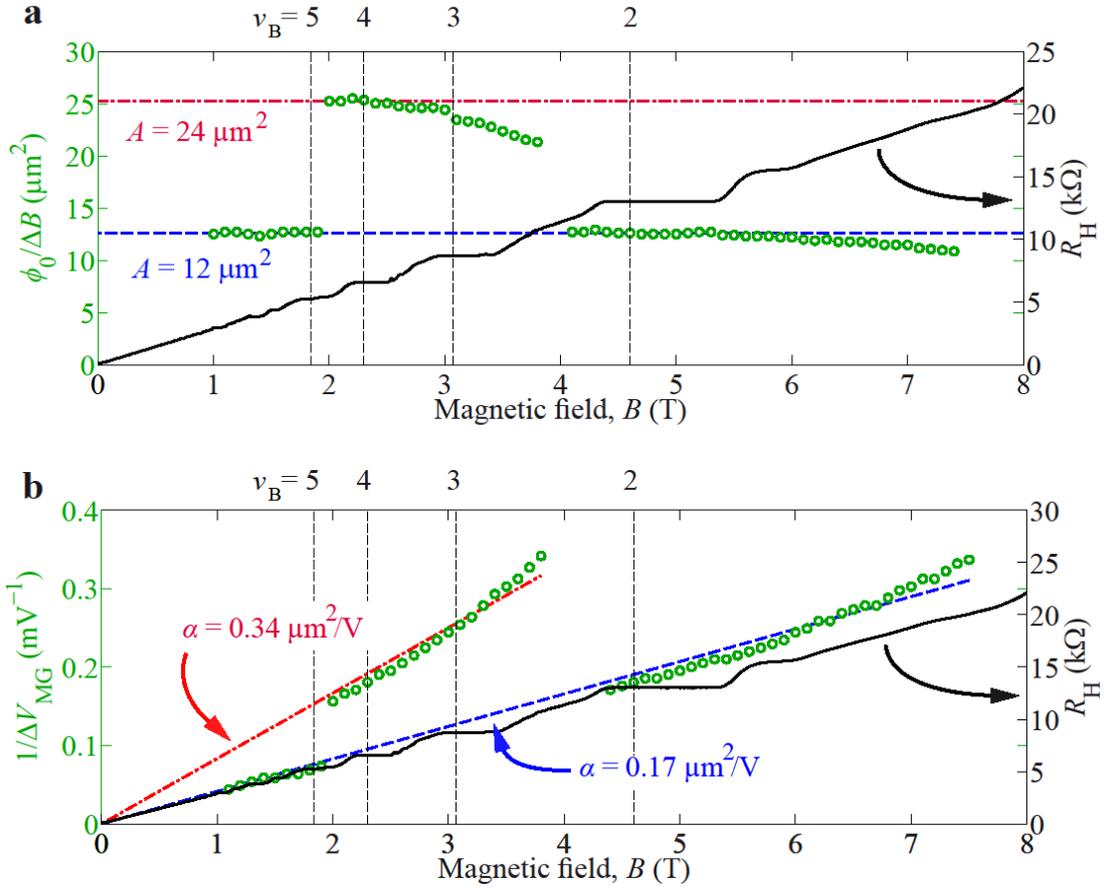

**Figure 3 | Aharonov-Bohm frequency as function of the magnetic field measured with a 12μm² FPI. a**, Frequency dependence on magnetic field (green, circles) alongside with bulk Hall resistance (black, solid line). **b**, Frequency dependence on modulation-gate voltage. In the two graphs, the regions $v_B<2.5$ and $v_B>4.5$ are in agreement with the expected AB oscillations. The frequency in the intermediate region, covering filling $v_B=3-4$, is doubled in $1/\Delta B$ and has a doubled slope in $1/\Delta V_{MG}$. This is attributed to a modified magnetic flux quantum $\phi_0=h/2e$ and quasi-particle charge $e^*/e=2$.

**Figure 4**

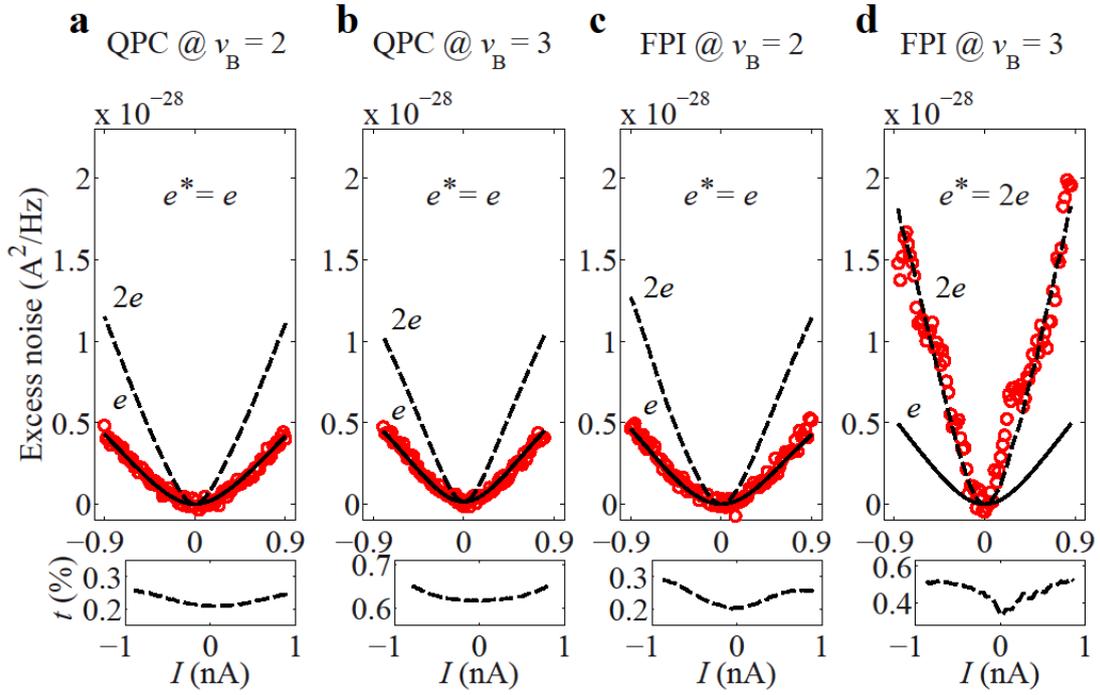

**Figure 4 | Excess shot noise and transmission revealing a quasi-particle charge in the different filling regimes**. Measured excess noise (red circles) and expected noise according to Eq. 2 for quasi-particle charge $e^*=e$ (black, solid) and for $e^*=2e$ (black, dashed) are plotted as function of source-drain DC current $I_{SD}$. **a,** Excess noise of a single QPC at $\nu_B=2$; **b**, Excess noise of a single QPC in $\nu_B=3$; **c,** Excess noise of a FPI at $\nu_B=2$; **d,** Excess noise of a FPI at $\nu_B=3$.

**Figure 5**

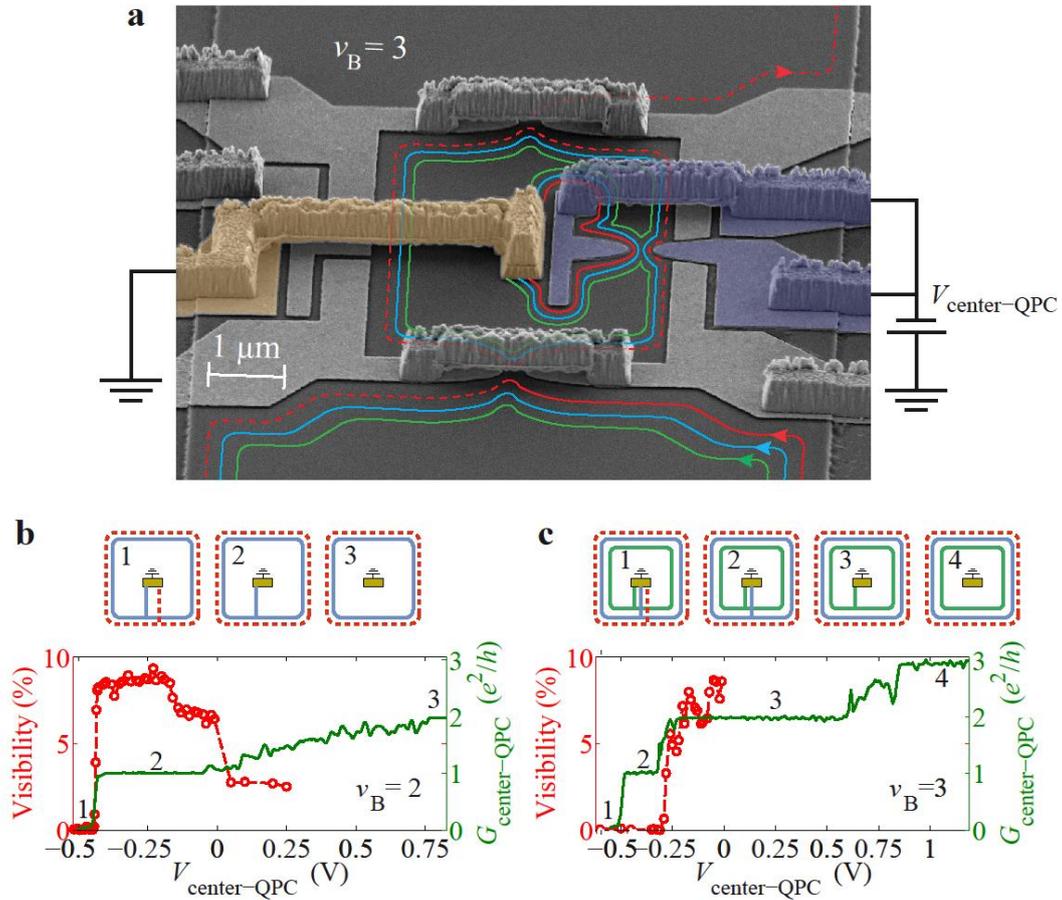

**Figure 5 | The effect of selective reflection of edge channels into the grounded center-ohmic contact on the visibility. a**, SEM image of a 12μm$^2$ FPI with a center-ohmic contact (gold) and an additional center-QPC (blue) placed along the FPI edge. An illustration of the edge channels is given for $v_B$~3 with the most inner edge channel reflected by the center-QPC into the center-ohmic contact. 'Cold' edges, originating from the ground, are not plotted. **b**, Conductance $G$ (green, solid line) of the center-QPC plotted alongside with the visibility of the AB oscillations (red, dashed line) at $v_B$=2. The visibility diminishes as the center-QPC reflects the outer channel to ground. **c,** The same measurement as in **b,** but at $v_B$=3. Surprisingly, here, the visibility fully diminishes once the second channel is reflected to ground.

**Figure 6**

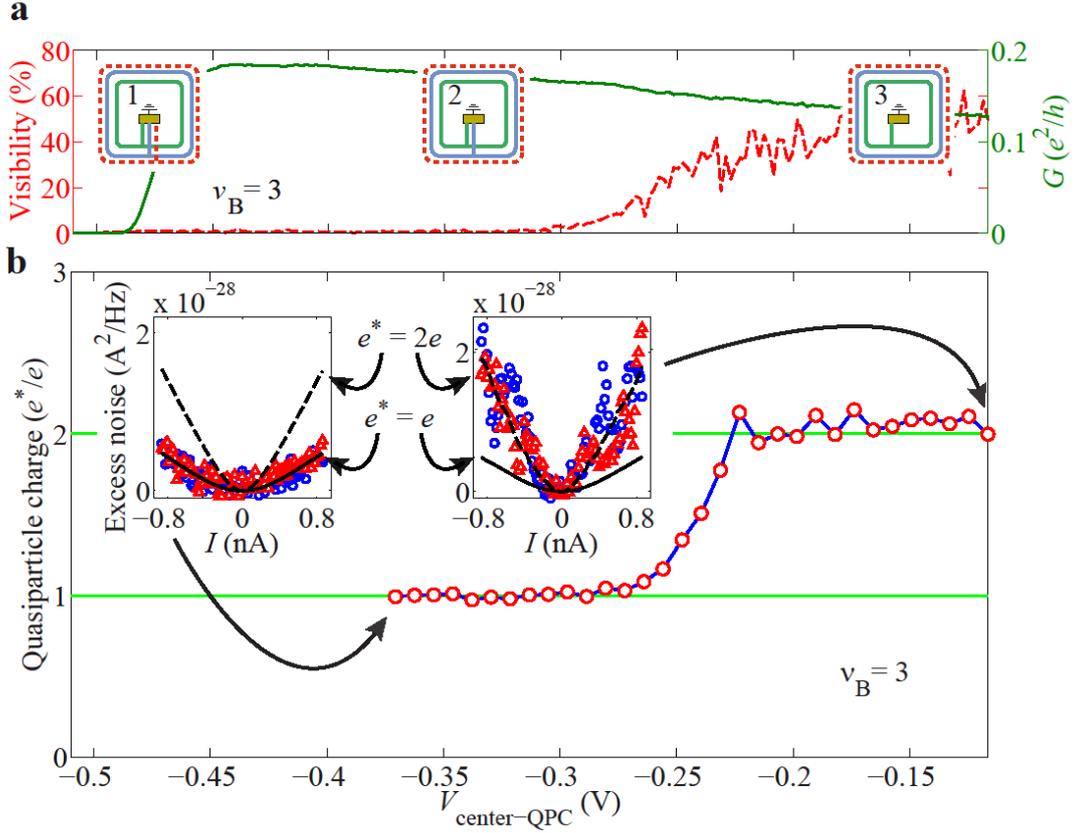

**Figure 6 | The effect of selective reflection of edge channels into the center-ohmic contact on the quasi-particle charge at the *h/2e* regime. a**, Conductance *G* of the FPI (green, solid line) and its visibility (red, dashed line) versus the center-QPC voltage $V_{center-QPC}$. An expanded scale of the conductance vanishing when the most outer edge channel is reflected to ground at the center-QPC, while the visibility diminishes when the second edge channel is reflected. **b**, The quasi-particle charge versus $V_{center-QPC}$. A clear coincidence between the diminishing of the visibility in **a** and the reduction of the charge from $e^* \sim 2e$ to $e^* = e$ is found. **Insets:** two characteristic noise measurements (red triangles and blue circles) from which charge is extracted and then averaged. The expected noise for the first is plotted according to Eq. 2 for $e^* = e$ (black, solid) and $e^* = 2e$ (black, dashed).

# Electron Pairing in the Integer Quantum Hall Effect Regime – Supplementary materials

H. K. Choi[1*], I. Sivan[1*], A. Rosenblatt[1], M. Heiblum[1], V. Umansky[1], D. Mahalu[1]

## 1. Density dependence

The reported phenomenon have been observed on materials with densities ranging 1.3-2.4×$10^{11}$cm$^{-2}$; demonstrating that the transition regions between the two periodicities ($h/e$ and $h/2e$) is an explicit function of bulk-filling-factor $\nu_B=n_e/n_B$, where $n_B$ is the density of flux piercing the interferometer. In order to further clarify this point we have made use of a top-gate, covering the whole area of the FPI, which allowed control the electronic density within the 2DEG, and as a result the filling-factor within the FPI $\nu_{FPI}$.

In Fig. S1 we show (blue) the conductance of the FPI as function of the top-gate voltage, while the QPCs are unbiased; namely, the QPCs are fully open and the conductance is governed by the density below the top-gate. Then, via the top-gate voltage, we set the bulk-filling within the FPI to $\nu_{FPI}$=6 (green, circle) or $\nu_{FPI}$=4 (red, triangle). In the two insets, we show the FFTs of AB oscillations (of the most outer edge channel) measured at these filling factors. As anticipated according to the results shown in the main text (see Fig. 3), at $\nu_{FPI}$=6 we observe a period corresponding to an effective area $\phi_0/\Delta B^e$=12.5μm$^2$, while at $\nu_{FPI}=4$ the effective area doubles $\phi_0/\Delta B^{2e}$=25μm$^2$.

## 2. Possibility of even windings

A striking point is the relative values of the highest visibilities in the two regimes, being similar and around 50% in the larger devices of size 12μm$^2$. Consequently, the $h/2e$ periodicity cannot be attributed to even-winding process. As we demonstrate below, with an estimated dephasing length, the maximal visibility of a two-winding interference cannot exceed 10%.

The amplitude of traversing the FPI is given, up to a phase factor, by $\tau = t^2 \sum_{n=0}^{\infty}(r^2 e^{i\phi_{AB}})^n$, where we took $r_l = r_r$ and $t_l = t_r$. Evidently, this sum is given by $\tau = \frac{T}{1-Re^{i\phi_{AB}}}$, where $T = t^2$ and $R = r^2 = 1 - T$ are the transmission and reflection probabilities respectively. It is not straightforward how to make approximations regarding the coherence length of the interferometer using this expression; for example, no approximation on $R$ can be made if we are interested in both high and low transmission regimes. On the other hand, we may consider approximations on the number of coherent windings that can be undertaken in the FPI. For this purpose we can write the probability of traversing the FPI as $|\tau|^2 = T^2 \sum_{n,m=0}^{\infty} (R \cdot e^{i\phi_{AB}})^n (R \cdot e^{-i\phi_{AB}})^m$, or equivalently as:

$$|\tau|^2 = T^2 \sum_{s,p} R^s \cdot e^{ip\phi_{AB}},$$

Where $s=n+m$ is the total number of windings undertaken in the two interfering paths, and $p=n-m$ is their phase difference, or the harmonic. In order to better understand this we write these in the form of matrices:

$$p = \begin{pmatrix} 0 & 1 & 2 & 3 & 4 & \\ -1 & 0 & 1 & 2 & 3 & \cdots \\ -2 & -1 & 0 & 1 & 2 & \\ -3 & -2 & -1 & 0 & 1 & \\ -4 & -3 & -2 & 1 & 0 & \\ \vdots & & & & & \ddots \end{pmatrix}$$

$$s = \begin{pmatrix} 0 & 1 & 2 & 3 & 4 & \\ 1 & 2 & 3 & 4 & 5 & \cdots \\ 2 & 3 & 4 & 5 & 6 & \\ 3 & 4 & 5 & 6 & 7 & \\ 4 & 5 & 6 & 7 & 8 & \\ \vdots & & & & & \ddots \end{pmatrix}$$

In the $p$ matrix, the main diagonal ($n=m \rightarrow p=0$) represents the DC conductance, having no AB-phase factor. Then, each pair of diagonal lines above and below it ($m=n\pm i \rightarrow p=\pm i$) stands for the $i$'th harmonic. For example, the $p=\pm 1$ represents the interference between two paths that differ by a single winding. Now the phenomenological approximation that we take (for a FPI size $A=12\mu m^2$) is that of having no more than two coherent windings, which reads:

$$p = 0; s = 0,2,4, \dots$$
$$p = \pm 1; \ s = 1$$
$$p = \pm 2; \ s = 2,$$

denoted in red, blue and green in the matrices. Thus we obtain the following expression:

$$|\tau|^2 = T^2 \left[ \frac{1}{1-R^2} + 2R \cdot \cos\phi_{AB} + 2R^2 \cdot \cos 2\phi_{AB} \right].$$

Thus, we can deduce the following visibilities for the first and second harmonics:

$$\nu^{(1)} = 2(1-R^2)R$$

$$\nu^{(2)} = 2(1-R^2)R^2 \ .$$

In order to account for possible dephasing in the system we add a random-phase component to our transmission amplitude $\tau = t^2 \sum_{n=0}^{\infty} \left( r^2 e^{i\phi_{AB}+i\varphi_{rand}^{(n)}\gamma} \right)^n$, where $\varphi_{rand}^{(n)} \in [0,2\pi]$ is a random phase to be integrated and $\gamma \equiv L/L_\gamma \in [0,1]$ where $L$ is the electrons' path length, and $L_\gamma$ is the dephasing length, a property of the device, presumably a function of temperature and gate-stability. Under the same assumptions presented previously we now obtain that the transmission through the FPI reads:

$$(3) \quad |\tau|^2 = T^2 \left[ \frac{1}{1-R^2} + 2R \cdot \frac{\sin(\pi\gamma)}{\pi\gamma} \cdot \cos\phi_{AB} + 2R^2 \cdot \frac{\sin(2\pi\gamma)}{2\pi\gamma} \cdot \cos 2\phi_{AB} \right],$$

leading us to the following expression:

$$(4) \quad v^{(1)} = 2(1-R^2)R \frac{\sin(\pi\gamma)}{\pi\gamma}$$

$$(5) \quad v^{(2)} = 2(1-R^2)R^2 \frac{\sin(2\pi\gamma)}{2\pi\gamma}$$

Applying Eq. 4 to our experimental result of $v^{(e,1)}_{max} \approx 60\%$, where the index $(e,1)$ represents the 1st harmonic in the $h/e$ regime, we find $\gamma \approx 0.4$, setting an upper-bound for the maximal visibility of the second harmonic $v^{(e,2)}_{max} \approx 10\%$ (for a FPI size $A=12\mu m^2$). This upper-bound is one order of magnitude lower than the experimental result $v^{(2e,1)}_{max} \approx 50\%$, suggesting that the $h/2e$ periodicity is not due to the favoring of two windings paths.

### 3. Transmission dependence

We stress that no $h/e$ periodicity is observed at the $h/2e$ regime, and vice-versa (save for the transition regions). Moreover, the AB periodicity in these two regions has no dependence on the two QPCs' transmissions and not on the overall transmission, nor on the degree of asymmetry between the two QPCs. In Fig. S2 we show the conductance through a FPI as function of the two QPCs (a) at $v_B=2$, and (b) at $v_B=3$. In the insets we show the evolution of the interferences as the two QPCs are opened simultaneously (red to blue); with only a single frequency observed at all times.

The result shown in Fig. S2 is of considerable importance: if the appearance of $h/2e$ periodicity results from the self-capacitance of the interfering edge channel, one may assume that pinching the FPI will decrease that capacitance, and affect the paired periodicity – which was not found. As evidence show, the inter-channel mutual capacitance is likely to be important.

### 4. Interference of inner

Here we show that the $h/2e$ periodicity appears only when the most outer edge channel interferes. In Fig. S3 we show the conductance through a FPI as function of both QPCs (a) at $v_B=3$, and (b) at $v_B=4$. The blue circles (red triangles) denote partitioning of the second outer (most outer) edge channel at the QPCs. The corresponding AB periodicities and their FFTs are shown in the insets; demonstrating a clear 1:2 relation. Though, the number of inner channels in both cases is similar, yet, the AB periodicity of the second edge channel is $h/e$ (always).

### 5. Temperature dependence

The temperature dependence of the visibility in the two different regimes ($h/e$ and $h/2e$) is very similar, as shown in Fig. S4. In the $h/e$ regime, measured at $v_B=2$, the visibility decays exponentially as $e^{-T/T_0}$, with $T_0=40$mK; however, in the $h/2e$ regime, measured at $v_B=3$, the visibility decays similarly, with $T_0=25$mK. We do not attribute to this small difference any importance since it is on the order of the error.

**6. Coexistence of the two frequencies**

In the low-field narrow transition region around $v_B=4.5$, we find a region in which the two frequencies coexist. The phase-relation between the two frequencies, as well as the fashion in which the one evolves into the other might shed light on the reported phenomena. Fig. S5 shows a characteristic pajama measurement (as the ones shown in the main text, see Fig. 2) in the transition region. We find it to follow:

$$(6) \quad C_e \cos(2\pi \cdot AB/(h/e)) + C_{2e}\cos(2\pi \cdot AB/(h/2e)) \;,$$

where $C_e$ and $C_{2e}$ are the amplitudes of $h/e$ and $h/2e$ components respectively.

Although the two frequencies are correlated in phase, they appear to be independent in nature. In an experiment similar to that described in the main text (see Fig. 5), we show in Fig. S6 the transmission of the center-QPC (green, solid) as a function of the voltage applied to it $V_{\text{center-QPC}}$, alongside with the oscillation's visibility of the $h/e$ (blue circles) and $h/2e$ (red triangles). It clearly shows that when the second outer channel is reflected to ground via the center-QPC the $h/2e$ periodicity vanishes, whereas the $h/e$ periodicity remains and even gains visibility. This experiment proves the independence of the two periodicities, which may share some 'mutual coherence'.

**7. Different samples**

Here is stress the universality of the reported phenomena by measuring a different samples. As mentioned in the main text, only a few details seem to vary from one sample to another; such as the behavior around the transition regions.

Fig. S7 shows the magnetic field dependence of the AB frequency $1/\Delta B$ in a top-gated FPI of size $A=12.6\mu m^2$ (similar to that in Fig. 3a). Similarly, Fig. S8 shows the magnetic field dependence of the AB frequency $1/\Delta V_{\text{MG}}$ for a top-gated FPI of size $A=12.6\mu m^2$ (similar to Fig. 3b).

**8. Shot Noise**

As explained in the main text, we employ Eq. 2 using the overall transmission of the FPI, being a function of the AB phase and the two QPCs. In Fig. S9 we demonstrate that indeed, the well-known equation of shot-noise can be applied to the FPI, by a thorough measurement at $\nu_B=2$, where we expect the quasiparticle charge to be that of an electron. We trace the quasiparticle charge (Fig. S9a), the visibility (Fig. S9b) and conductance (Fig. S9c) of the FPI, as we scan the source-QPC from high to low transmission while keeping the drain-QPC fixed to partition the most outer channel. Noticeably, while the conductance and visibility vary substantially, the charge remains $e^*=e$.

## 9. Absence of tunneling between edge channels

In order to check whether tunneling is present in the $h/2e$ regime the experiment depicted in Fig. S10 was performed. While the most outer edge channel interferes at $\nu_B=3$ (with $h/2e$ oscillations), the center-QPC is being gradually pinched ($V_{center-QPC}$ in the Fig. S10), reflecting edge channels selectively to the contact. The current at each reflected channel to the center contact is being probed by the voltage across a 1kΩ resistor connecting the center contact to ground.

The current flowing through the FPI (green, solid line) decreases abruptly at the transition between regions 2 and 1 (left arrow), which is the point in which the most outer edge channel is reflected at the center-QPC. Naturally, at this point the current flowing into the center ohmic contact (green, dashed line) rises. Both currents decrease linearly due to the unavoidable electrostatic coupling between the center-QPC and the FPI's QPCs. We notice that the visibility decreases abruptly at the transition between regions 3 and 2 (right arrow), which is the point in which the second outer edge is reflected at the center-QPC.

If tunneling were present between the most outer and second outer edge channels, then it would result in current into the center ohmic contact in region 2 and/or at the transition between 3 and 2. Namely, in this region, if tunneling were present, then electrons from the most outer edge channel would hop into the second outer one, and from there to the ground, resulting in a reduction of the current flowing to the drain and an increase in the current flowing into the small contact. Nonetheless, no such signature of tunneling current was observed.

**Figure S1**

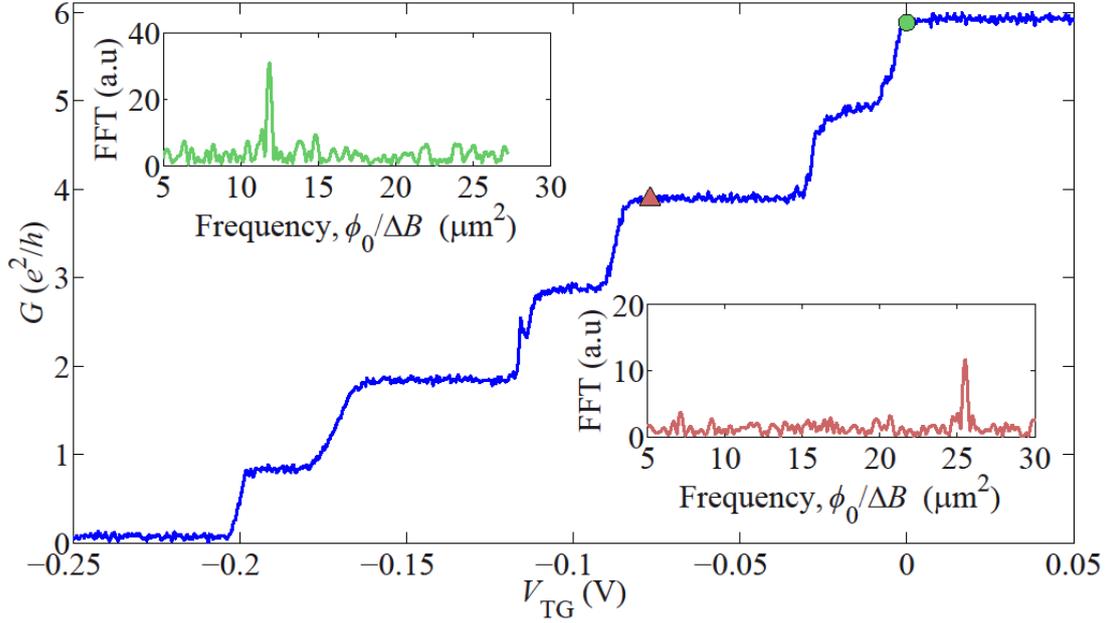

**Figure S1 | Demonstrating a passage between *h/e* and *h/2e* induced by varying the filling factor via the top-gate voltage.** The conductance $G$ through the FPI versus the top-gate voltage $V_{TG}$, determining the filling factor within the FPI $\nu_{FPI}$ via its electronic density $n_e$, according to $\nu_{FPI}= n_e/ n_B$, where $n_B$ is the density of flux piercing the interferometer. Insets: FFTs of the AB oscillations at two values of the top-gate voltage corresponding to effective $\nu_{FPI}=6$ and $\nu_{FPI}=4$. We see that the resonance frequency in these two filling factors has a factor two, coinciding with the observation shown in Fig. 3 in the main text.

**Figure S2**

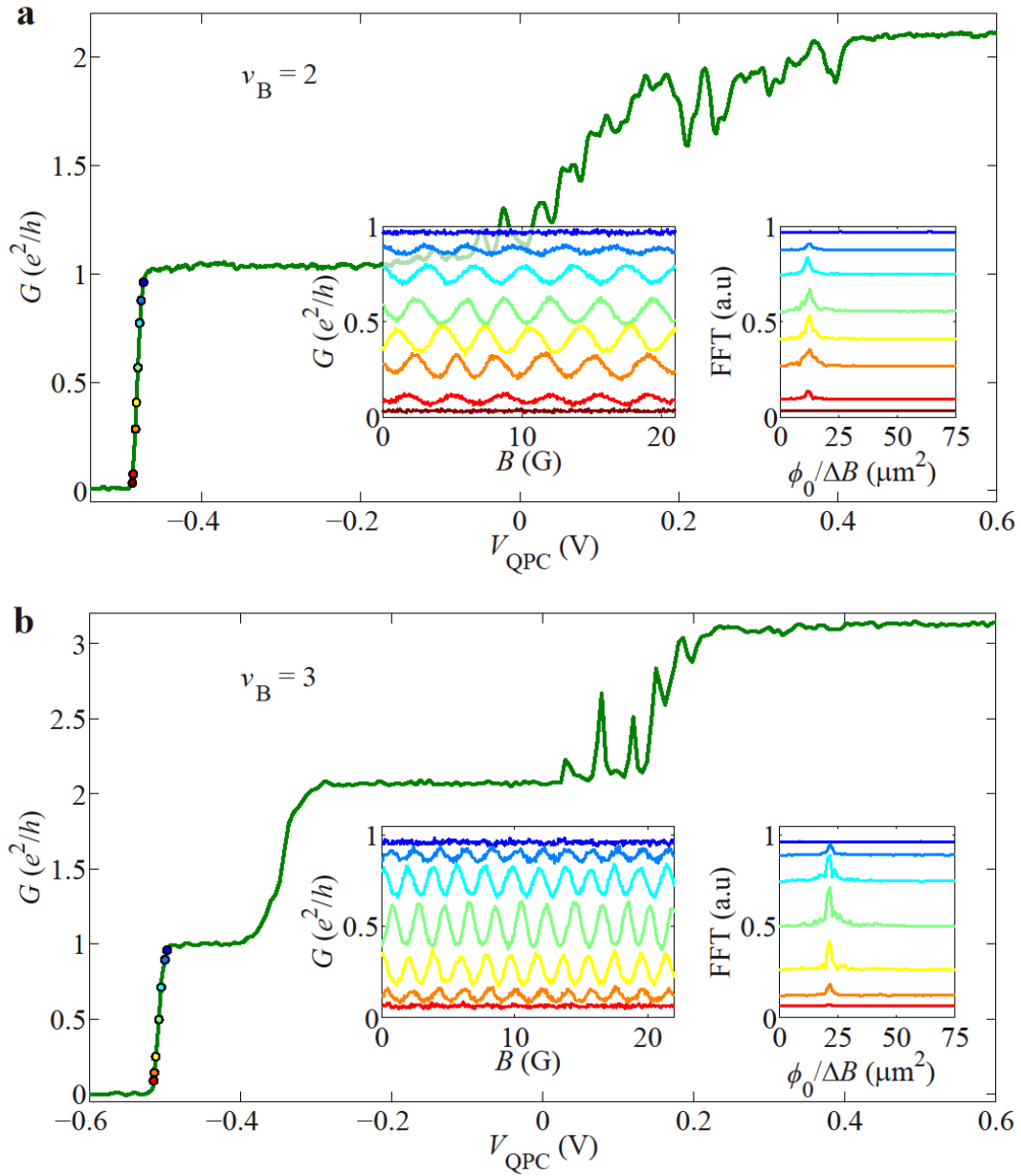

**Figure S2 | Demonstrating that the frequency of the AB oscillations do not depend on the degree of opening of the FPI's QPCs.** Conductance $G$ through the FPI as both its QPC are biased together. (Inset) Left: Evolution of the AB oscillations of the most outer edge channel as the FPIs' QPCs are being closed at the $h/e$ (**a**) and $h/2e$ (**b**) regimes. Right: Their corresponding FFT. These measurements point out that when $h/2e$ oscillations govern the oscillations, no transition to $h/e$ occurs, as long as the filling factor remains.

Figure S3

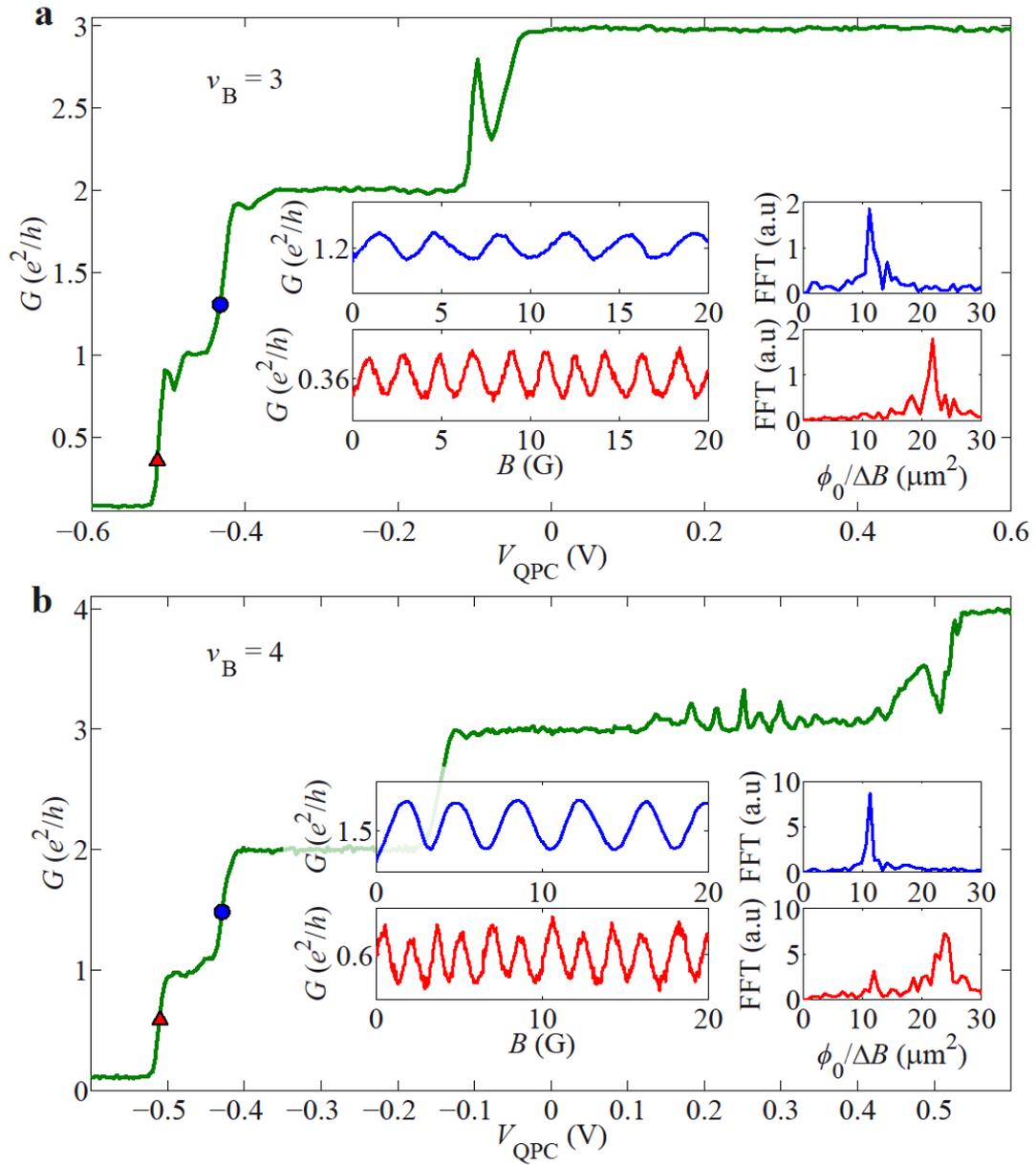

**Figure S3 | Difference in AB frequency between the most outer and second outer edge channels.** The conductance $G$ through the FPI as a function of both FPIs' QPCs voltage $V_{QPC}$ at $v_B=3$ (**a**) and $v_B=4.5$ (**b**). Insets: AB oscillations and their FFTs at two values of $V_{QPC}$ in which the most outer channel is partitioned (red triangle) and the second outer is partitioned (blue circle).

**Figure S4**

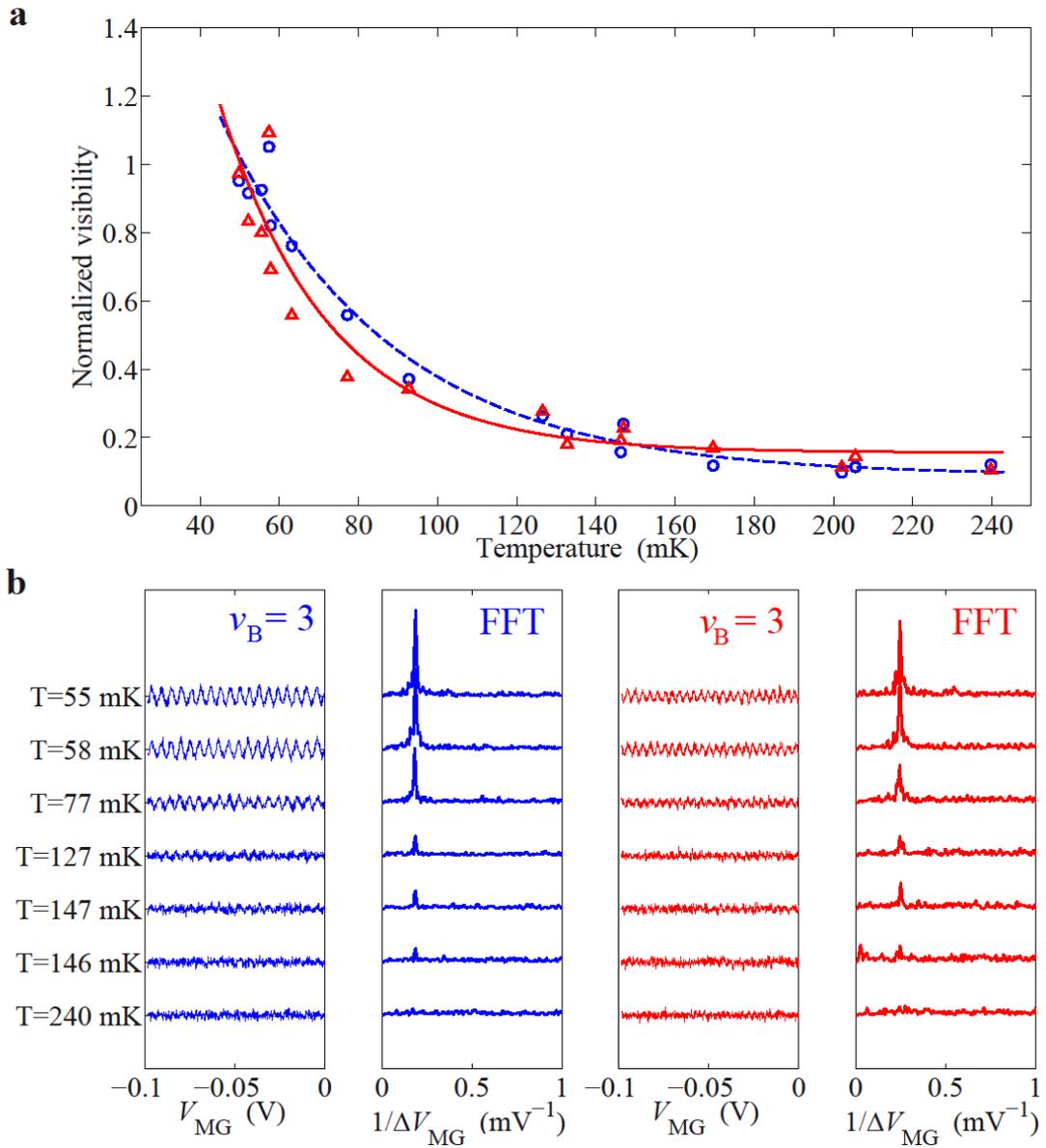

**Figure S4 | Effect of temperature on the two types of oscillations. a**, Visibility of $h/e$ AB oscillations (Measurement: blue circles, fit: dashed blue) measured at $\nu_B=2$ and $h/2e$ (Measurement: red triangles, fit: solid red) measured at $\nu_B=3$. In both cases a similar decay rate is observed. **b**, AB oscillations versus the modulation-gate voltage $V_{MG}$, and the corresponding FFT, at different temperatures.

**Figure S5**

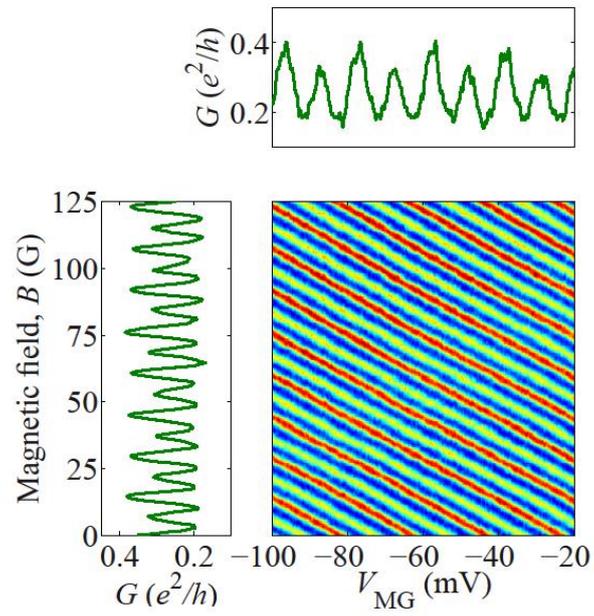

**Figure S5 | Pajama measurement in the transition region.** Color-plot of the conductance *G* of a FPI with respect to magnetic field and modulation-gate voltage $V_{MG}$, similar to the ones shown in Fig. 2.

**Figure S6**

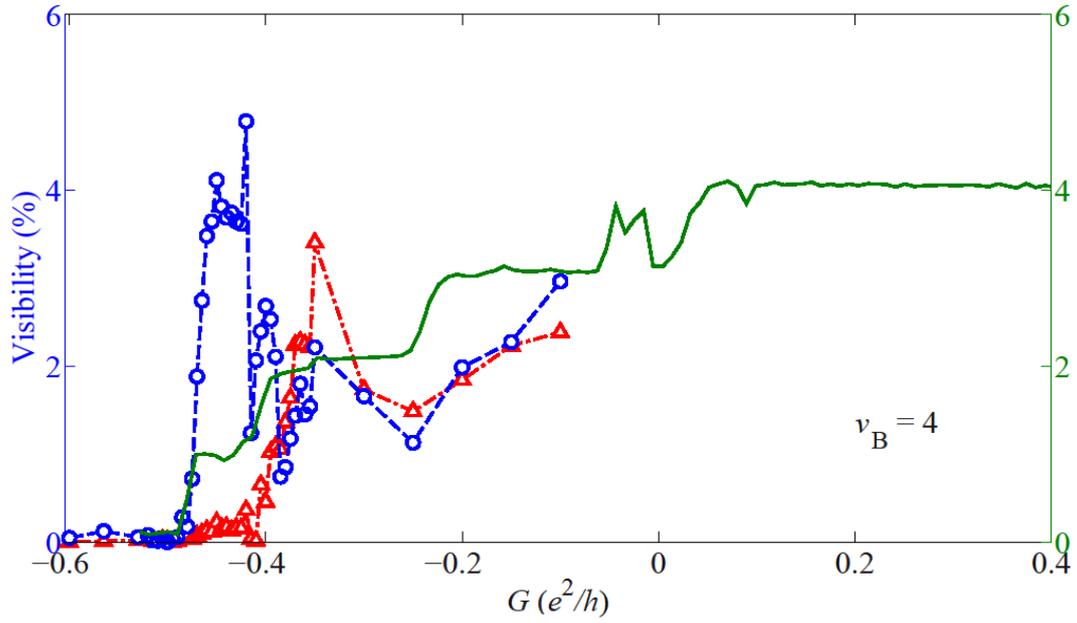

**Figure S6 | Dephasing the *h/2e* oscillations in the transition region does not affect the coherence of the *h/e*.** Conductance $G$ (green, solid) through the center-QPC alongside with the visibility of the two interference frequencies, *h/e* (blue circles) and *h/2e* (red triangles). We vary the number of channels which are reflected into the center contact, as illustrated in Fig. 5 (main text). Surprisingly, once the second outer channel is dumped into the center-contact, the visibility of the *h/2e* oscillations fully diminishes to zero, whereas the *h/e* visibility grows higher.

Figure S7

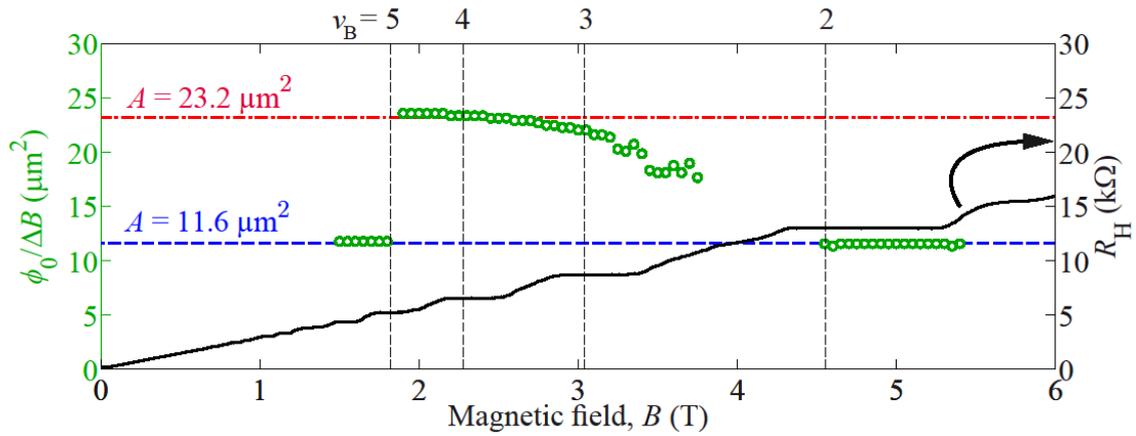

**Figure S7 | Aharonov-Bohm frequencies as function of the magnetic field measured with a 11.6µm² top-gated FPI**. Frequency in magnetic field (green, circles) and the Hall resistance of the sample (black, solid line).

Figure S8

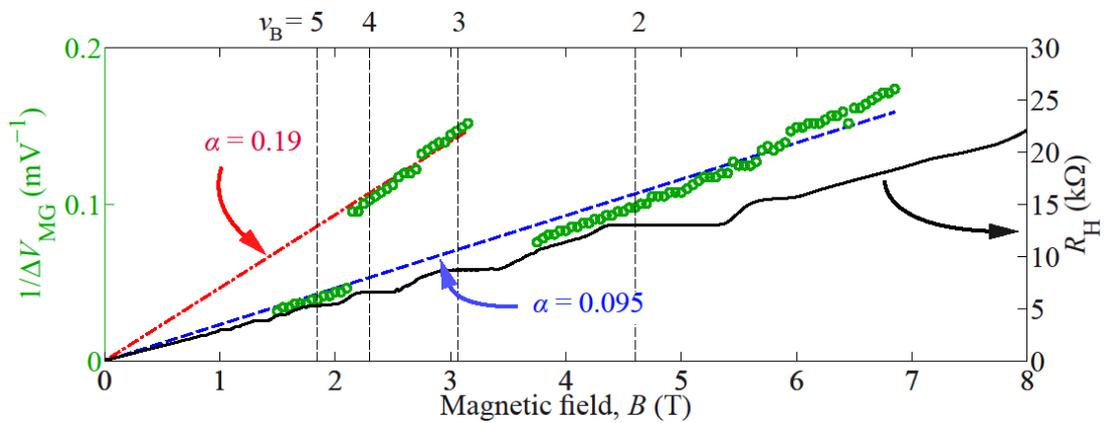

**Figure S8 | Aharonov-Bohm frequencies as function of the magnetic field measured with a 2µm² FPI with grounded ohmic contact**. Frequency in MG voltage (green, circles) and the Hall resistance of the sample (black, solid line).

**Figure S9**

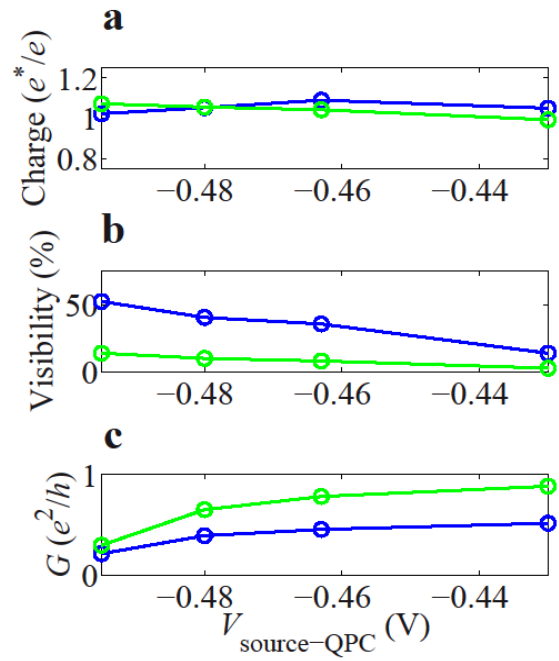

**Figure S9 | Demonstrating that the FPI at $\nu_B=2$ has quasiparticle charge $e^*=e$, as a single QPC.** – will be redone --Effective charge, visibility, and conductance $G$ as function of the drain-QPC. The two colors stand for two settings of QPC-left, both partitioning the LLL. On the right side our device is a single QPC, while on the left side the device is a FPI. Noticeably, the effective charge remains approximately constant, around the value of the electron charge.

**Figure S10**

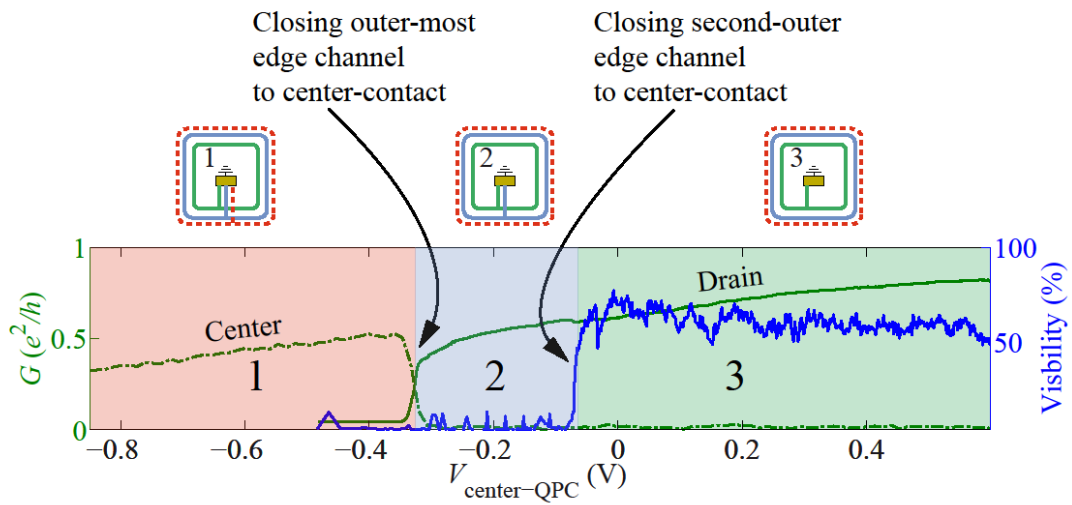

**Figure S10 | Demonstrating that no tunnel was observed between the most outer and second outer edge channels.** Conductance between source and drain through the FPI (green, solid), conductance between source and center contact (green, dashed) and the visibility of the AB interference of the most outer edge channel (blue). The transitions between the different regimes (denoted in colors, with an illustration above) correspond to the abrupt reflection of an edge channel at the center-QPC. As explained in the text, observation that the current does not abruptly increase at the left transition demonstrate that no tunneling between the edges is present.